\documentclass[mathleft]{an}
\usepackage{graphicx}
\usepackage{times}
\begin{document}
\Pagespan{383}{} 
\Yearpublication{2006}%
\Yearsubmission{2005}%
\Month{4}%
\Volume{327}%
\Issue{4}%
\DOI{10.1002/asna.200510538}%

\title{Concerning the instantaneous mass and the extent of an expanding\\ universe}
\author{H.J. Fahr\inst{1}\fnmsep\thanks{Corresponding author: hfahr@astro.uni-bonn.de}
\and M. Heyl\inst{2}}

\institute{ Institute for Astrophysics and Extraterrestrial
Research (now part of Argelander-Institut f\"{u}r Astronomie),
University of Bonn, Auf dem H\"ugel 71, 53121 Bonn, Germany
\and Deutsches Zentrum f\"{u}r Luft und Raumfahrt (DLR),
K\"{o}nigswinterer Stra{\ss}e 522, 53227 Bonn, Germany}

\date{Received 2005 Oct 6}
\accepted{2006 Jan 24}
\publonline{2006 Mar 30}

\abstract{In this article we want to answer the cosmologically
relevant question what, with some good semantic and physical reason,
could be called the mass $M_\mathrm{u}$ of an infinitely extended,
homogeneously matter-filled and expanding universe. To answer this
question we produce a space-like sum of instantaneous cosmic energy
depositions surrounding equally each spacepoint in the homogeneous
universe. We calculate the added-up instantaneous cosmic energy per
volume around an arbitrary space point in the expanding universe. To
carry out this sum we use as basic metrics an analogy to the inner
Schwarzschild metric applied to stars, but this time applied to the
spacepoint-related universe. It is then shown that this leads to the
added-up proper energy within a sphere of  a finite outer critical
radius defining the point-related infinity. As a surprise this
radius turns out to be reciprocal to the square root of the
prevailing average cosmic energy density. The equivalent mass of the
universe can then also be calculated and, by the expression which is
obtained here, shows a scaling with this critical radius of this
universe, a virtue of the universe which was already often called
for in earlier works by E.Mach, H.Thirring and F.Hoyle and others.
This radius on the other hand can be shown to be nearly equal to the
Schwarzschild radius of the so-defined mass $M_\mathrm{u}$ of the universe. }

\maketitle

\section{Introduction to a mass definition of the universe}

The question already often has been raised by cosmologists (see e.g. Tolman, 
1918, Thirring, 1918, Jordan, 1947, Mashhoon, 1988, Mashhoon et al., 1984, Rosen
\& Copperstock, 1992, Barbour, 1995) how the so-called mass $M_\mathrm{u}$ of an
expanding universe, filled with homogeneously deposited matter subject to a
Hubble flow, should semantically reasonably and physically adequately be
conceived and what as such it may represent in form of a quantifiable cosmic
number or a cosmic reality. How should this mass $M_\mathrm{u}$ be related to
cosmological ingredients which are closer associated to the observable
universe like the mean mass density, the cosmic scale variation with time or
the Hubble parameter, or the so-called $\Omega-$ value, relating cosmic mass
density and the critical density $\rho_\mathrm{crit}$? Especially in an isotropically
expanding, homogeneous universe with a vanishing or negative curvature
parameter $k\leq0$ in which the volume of the universe is infinite, the
definition of such a quantity like $M_\mathrm{u}$ at first glance does not seem to be
straightforward, perhaps not making sense at all.

Starting with suggestions made by Dirac (1937) and investigations carried out
by Schuecking (1954) and Einstein \& Straus (1945) ideas were developing and
concretizing that the masses of elementary particles as well as of stars might
be somehow related to the large scale structure of the universe and its
development in cosmic time. Much earlier already Mach (1883) had formulated as
a physical requirement that the inertial mass of each particle just by its
rigorous concept should somehow be related to the constellation of all other
masses in the universe. Though Mach`s principle never up to the present
completely entered into theoretical formulations of mass and gravity
interacting with eachother, this principle already pointed to the probable
fact that masses and their constellation in the universe should be related to
eachother. This idea was later taken up by Thirring 1918 who was asking the
question what the absolute inertial rest frame would be in the presence of
rotating masses. In his view it should lead to identical geoidic deformations
of the rotating earth regardless whether the earth rotates with respect to the
universe with an angular frequency $\omega,$ or whether the universe rotates
with $-\omega$ with respect to the earth at rest.

To put these two cases on a physically conceivable basis, he compared two
situations: a) the earth rotating with $\omega$ in the center of a universe at
rest represented by a massive spherical shell with radius $R_\mathrm{u}$ and mass
$M_\mathrm{u}$, and b) the earth at rest in the center of the universe represented
identical to case a), however, this time rotating with $-\omega$ with respect
to the earth. The result of this comparison, when extracted from the adequate
Newtonian approximation of Einstein`s field equations for rotating masses,
Thirring (1918) (see also Mashhoon et al. 1984) found that at a space point
close to the axis in the rotation plane the centrifugal or geodetic forces
operating at the earth's equator in cases a) and b) are related by

\begin{equation}
\textbf{K}_\mathrm{centri}=X\cdot\textbf{K}_\mathrm{geod} , \label{1}%
\end{equation}
where the factor $X$ is calculated to be:
\begin{equation}
X=\frac{4GM_\mathrm{u}}{3c^{2}R_\mathrm{u}}=\frac{2}{3}\frac{R_\mathrm{s,u}}{R_\mathrm{u}} , \label{2}%
\end{equation}
with the associated Schwarzschild radius $R_\mathrm{s,u} = 2GM_\mathrm{u}
c^{-2}$ of the mass $M_\mathrm{u}$. This expresses the fact that the
equivalence principle of rotational motions can only be satisfied in the
present and upcoming world represented by values $M_\mathrm{u}$ and $R_\mathrm{u}$, if the
value for $X$ would exactly evaluate to be constant and equal to:\textbf{
}$X=1$.\textbf{ }To preserve $X=1$ in an expanding universe would, however,
require the expression $M_\mathrm{u}/R_\mathrm{u}$ to be a cosmological constant\textbf{.
}This would mean that the total mass of the universe should scale with the
increasing cosmic diameter $R_\mathrm{u}$, an unusual cosmological requirement.

In the following we shall not rely on the above indicated scaling of the mass
of the universe with cosmic extent, we rather shall ask the more principle
question, how mass $M_\mathrm{u}$ and extent $R_\mathrm{u}$ of an expanding universe should
be properly defined. Hereby we simply rely on the cosmological principle
according to which each space point in the universe is equivalent to any other
cosmic space point concerning its being surrounded by matter-filled cosmic
space, and shall ask by how much instantaneously countable matter, i.e. matter
distributed on a 3-dimensional spacelike hypersphere of the universe, this
selected point of the universe may in fact be surrounded. This specific
question in our perspectives here is not connected with the problem of how
long it may take, before the presence of distant masses in the universe can be
communicated by electromagnetic or gravitational waves to our local standpoint
in the universe, i.e. we do not intend to sum up retarded contributions of
energy depositions in space. The question rather starts from the certainty and
knowledge that any spacepoint in the universe is surrounded up to large cosmic
distances by cosmic matter that is subject to a cosmological expansion
dynamics. What - then - may be called the instantaneous energy content of this
spacepoint-related universe?

\section{Calculation of the proper radius}
The universe, taken as basis of our calculations, is assumed to be expanding
isotropically and to have an isotropic curvature and a homogeneous energy
distribution. As the total instantaneous energy of such a universe considered
as an isolated system Tolman (1934) has given the following formula for the
mass $M_\mathrm{u}$ of such a system:%
\begin{equation}
M_\mathrm{u}(t)c^{2}=\int^{3}[\rho(t)c^{2}+3p(t)]\sqrt{-g_\mathrm{3}}\mathrm{d}^{3}V , \label{3}%
\end{equation}
where $\rho(t)$ and $p(t)$ denote mass density and pressure of the cosmic
matter, and where $\mathrm{d}^{3}V_\mathrm{0}=\sqrt{-g_\mathrm{3}}\mathrm{d}^{3}V$ denotes the differential of
the spacelike 3-D proper volume.

In later phases of an evolving and expanding universe rather than pressure
(n.b. matter becomes pressure-free!) as forms of positive energy
representations, in addition to rest mass energy distributions of baryonic or
dark matter type, also the kinetic or dynamic energy connected with the
cosmological expansion would need to be considered. In an homologously
expanding universe seen from any arbitrary spacepoint matter comoving with the
cosmic expansion in its surroundings shows up with a Hubble flow. As judged
from this spacepoint comoving matter at a distance $r$ thus also represents
kinetic energy in form of its Hubble motion with a centripetal velocity
$v_\mathrm{c}(r)=H(t)\cdot r$. Thus for a point-related energy balance account this
form of Hubble-induced kinetic energy needs to be counted in addition to rest
mass energy. To calculate the point-related total energy balance one may
describe the surroundings of an arbitrary spacepoint, similar to the central
spacepoint in the stellar interior, as surrounded by metrically relevant,
space-curving energy. For such point-related balance the metric tensor
elements of the so-called inner Schwarzschild metric tensor for the
point-related cosmic matter distribution can be applied.

Therefore adding up the above mentioned energies in the whole universe filled
with a pressure-less matter surrounding an arbitrary space point up to a
distance $R_\mathrm{u}$ may then lead one instead of Equ. (\ref{3}) to the following expression:%
\begin{equation}
M_\mathrm{u}(t,R_\mathrm{u})c^{2}=\int^{3}[\gamma(r)\rho_\mathrm{0}(t)c^{2}]\sqrt{-g_\mathrm{3}}\mathrm{d}^{3}V
, \label{4}%
\end{equation}
where $\rho_\mathrm{0}(t)$ denotes the homogeneous rest mass density which is
variable only with cosmic time $t$ , but not with space coordinates, where
$\gamma(r)^{-2}=1-(v_\mathrm{c}(r)/c)^{2}$ is the Lorentz factor evaluated for
the local expansion velocity $v_\mathrm{c}(r)=H(t)r$ with $H(t)=v_\mathrm{c}(r)/r=(1/R_\mathrm{u})\mathrm{d}R_\mathrm{u}/\mathrm{d}t$, and where $\mathrm{d}^{3}V_\mathrm{0}=\sqrt{-g_\mathrm{3}}\mathrm{d}^{3}V$ is the
local differential spacelike proper volume of space given through the
determinant of the 3-D part of the inner Schwarzschild-metric tensor which
yields the worldline element $\mathrm{d}s$ in the form (see e.g. Stephani 1988):%
\begin{eqnarray}
\mathrm{d}s^{2}=\exp(\lambda(r))\mathrm{d}r^{2}&+&r^{2}[\mathrm{d}\vartheta^{2}+\sin^{2}\vartheta
\mathrm{d}\varphi^{2}]-\cr
&-&\exp(\nu(r))c^{2}\mathrm{d}t^{2}
 , \label{5}%
\end{eqnarray}
and the proper volume in the form:%
\begin{eqnarray}
\mathrm{d}^{3}V_\mathrm{0}&=& \sqrt{-g_\mathrm{rr}g_\mathrm{\vartheta\vartheta}g_\mathrm{\varphi\varphi}
}\mathrm{d}r\mathrm{d}\vartheta \mathrm{d}\varphi=\cr 
&&\ \ \ \ \ \ \ \ \ \ \ \ \ \ \ \ \  =\sqrt{\exp(\lambda(r)r^{4}\sin^{2}\vartheta}\mathrm{d}r\mathrm{d}\vartheta \mathrm{d}\varphi.
\label{6}%
\end{eqnarray}
The Robertson-Walker metric usually taken to describe the spacetime metrics in
a homogeneous universe in this specific case of a space-related view has been
replaced in view of the specific question posed here by the inner
Schwarzschild metric which here is used up to the metric infinity with no need
for a smooth connection to the Robertson-Walker metric. Then from Equ.
(\ref{4}) one obtains the total mass of the expanding universe with
pressure-less matter and homogeneous density distribution in the form:%
\begin{equation}
M_\mathrm{u}(t)c^{2}=4\pi\rho_\mathrm{0}(t)c^{2}\int_\mathrm{0}^{R_\mathrm{u}}\frac{\exp(\lambda
(r)/2)r^{2}\mathrm{d}r}{\sqrt{1-(\frac{Hr}{c})^{2}}} . \label{7}%
\end{equation}
Here the full energy density of cosmologically moving matter at r as seen from
an arbitrary spacepoint at $r=0$ is given by $\epsilon=\gamma(r)\rho_\mathrm{0}c^{2}$
where $\gamma(r)^{-2}=1-(Hr/c)^{2}$ is the local Lorentz factor
of the source point matter. The metric function $\exp[\lambda(r)]$ for the
inner Schwarzschild metric of a pressure-less matter universe, using the
analogy to the stellar case and following Stephani (1988), is given by:%
\begin{equation}
\exp[-\lambda(r)]=1-\frac{8\pi G}{rc^{2}}\rho_\mathrm{0}\int_\mathrm{o}^{r}\gamma(x)x^{2}\mathrm{d}x ,
\label{8}%
\end{equation}
and thus allows to obtain the total mass of the universe in the following form:%
\begin{eqnarray}
&&M_\mathrm{u}(t)=4\pi\rho_\mathrm{0}(t)\times\cr 
&&\times\int_\mathrm{0}^{R_\mathrm{u}}\frac{r^{2}}{\sqrt{1-\frac{8\pi G}{rc^{2}}\rho
_\mathrm{0}\int_\mathrm{o}^{r}\gamma(x)x^{2}\mathrm{d}x}\sqrt{1-(\frac{Hr}{c})^{2}}}\mathrm{d}r . \label{9}%
\end{eqnarray}
As one can see from the above expression, a singularity appears in the
integrand at a sourcepoint distance where the line element $\mathrm{d}s_\mathrm{r}%
=\sqrt{g_\mathrm{rr}}\mathrm{d}r$ tends to infinity, i.e. at some outer critical radius
$R_\mathrm{u}$ given by the following implicit relation:%
\begin{equation}
\frac{c^{2}}{8\pi G\rho_\mathrm{0}}=\frac{1}{R_\mathrm{u}}\int_\mathrm{o}^{R_\mathrm{u}}\frac{1}%
{\sqrt{1-(\frac{Hx}{c})^{2}}}x^{2}\mathrm{d}x , \label{10}%
\end{equation}
and reminding that the above integral is solved by:%
\begin{eqnarray}
&&\int_\mathrm{o}^{R_\mathrm{u}}\frac{1}{\sqrt{1-(\frac{H x}{c})^{2}}}x^{2}\mathrm{d}x =
(\frac{c}{H})^{3}\int_\mathrm{0}^{\frac{HR_\mathrm{u}}{c}}\frac{\xi^{2}\mathrm{d}\xi}{\sqrt{1-\xi
^{2}}}=\cr
&&(\frac{c}{H})^{3}(\frac{1}{2}\arcsin\frac{HR_\mathrm{u}}{c}-\frac{HR_\mathrm{u}}%
{2c}\sqrt{1-(\frac{HR_\mathrm{u}}{c})^{2}} , \label{11}%
\end{eqnarray}
and going now to the horizon limit, where $c=HR_\mathrm{u}$ must be expected, would
then bring us to:%
\begin{equation}
R_\mathrm{u}^{2}\frac{\pi}{4}=\frac{c^{2}}{8\pi G\rho_\mathrm{0}} , \label{12}%
\end{equation}
or yielding:%
\begin{equation}
R_\mathrm{u}=\frac{1}{\pi}\sqrt{\frac{c^{2}}{2G\rho_\mathrm{0}}} .
\label{13}%
\end{equation}
This result can also be interpreted as saying that a scaling of the density
$\rho_\mathrm{0}$ with the co-defined critical radius $R_\mathrm{u}$ of the universe holds
according to the following relation:%
\begin{equation}
\rho_\mathrm{0}(R_\mathrm{u})=\frac{c^{2}}{2\pi^{2}GR_\mathrm{u}^{2}} . \label{14}%
\end{equation}
Without the kinetic energy of the Hubble flow taken into account we instead of
the above result would obtain from Equ. (\ref{10}) the following simplified expression:%
\begin{equation}
\frac{c^{2}}{8\pi G\rho_\mathrm{0}}=\frac{1}{R_\mathrm{u}}\int_\mathrm{o}^{R_\mathrm{u}}x^{2}\mathrm{d}x=\frac
{1}{3}R_\mathrm{u}^{2} , \label{15}%
\end{equation}
leading to the result very similar to that obtained in Equ. (\ref{14}),
namely given by:%
\begin{equation}
\rho_\mathrm{0}(R_\mathrm{u})=\frac{3c^{2}}{8\pi GR_\mathrm{u}^{2}}=\frac{3\pi}{4}(\frac{c^{2}%
}{2\pi^{2}GR_\mathrm{u}^{2}}) . \label{16}%
\end{equation}
The difference between Hubble flow taken and not taken into account only
amounts to a factor $(3\pi/4)$.

\section{Calculation of the proper mass}
First let us take up the results for the case that the Hubble kinetics is not
taken into account and calculate the proper mass with Equ. (\ref{9}) by:%
\begin{equation}
M_\mathrm{u}(t)=4\pi\rho_\mathrm{0}(t)\int_\mathrm{0}^{R_\mathrm{u}}\frac{r^{2}}{\sqrt{1-\frac{8\pi
G}{rc^{2}}\rho_\mathrm{0}\int_\mathrm{o}^{r}x^{2}\mathrm{d}x}}\mathrm{d}r . \label{17}%
\end{equation}
Using the result of Equ. (\ref{15}) for $R_\mathrm{u}$, one then obtains:%
\begin{eqnarray}
M_\mathrm{u}(t)&=&4\pi\rho_\mathrm{0}(t)R_\mathrm{u}^{3}\int_\mathrm{0}^{1}\frac{\xi^{2}}{\sqrt{1-\xi^{2}}%
}\mathrm{d}\xi=\cr
&&4\pi\rho_\mathrm{0}(t)R_\mathrm{U}^{3}\frac{\pi}{4} , \label{18}%
\end{eqnarray}
which with Equ. (\ref{16}) leads to the result:%
\begin{equation}
M_\mathrm{u}(t)=\pi^{2}\rho_\mathrm{0}(t)R_\mathrm{u}^{3}=\frac{3\pi^{2}c^{2}}{8\pi G}R_\mathrm{u} , 
\label{19}%
\end{equation}
which expresses the fact that the above defined mass of the universe scales
with the critical radius of the universe $R_\mathrm{u}$ defined by Equ. (\ref{4}).

Using now the above relation one finds that Thirring`s relation given by Equ.
(\ref{2}) with the above introduced quantities appears to be fulfilled, since
one obtains:%
\begin{equation}
X=\frac{4GM_\mathrm{u}}{3c^{2}R_\mathrm{u}}=\frac{4G}{3c^{2}R_\mathrm{u}}\frac{3\pi^{2}c^{2}}{8\pi
G}R_\mathrm{u}=\frac{\pi}{2} . \label{20}%
\end{equation}

Now we shall take also into account the Hubble kinematics. Therefore using
now, instead of the above, the radius of the universe given by Equ.
(\ref{13}) one may then also calculate for a universe with Hubble flow the
total mass of the universe from Equ. (\ref{9}) in the form:%
\begin{equation}
M_\mathrm{u}(t)=4\pi\rho_\mathrm{0}(t)\int_\mathrm{0}^{R_\mathrm{u}}\frac{r^{2}\mathrm{d}r}{\sqrt{1-f_\mathrm{1}%
(r)f_\mathrm{2}(r)}\sqrt{f_\mathrm{3}(r)}} , \label{21}%
\end{equation}
with%
\begin{equation}
f_\mathrm{1}(r)=\frac{4}{\pi rR_\mathrm{u}^{2}}(\frac{c}{H})^{3} , \label{22}%
\end{equation}%
\begin{equation}
f_\mathrm{2}(r)=\frac{1}{2}\arcsin\frac{Hr}{c}-\frac{Hr}{2c}\sqrt{1-(\frac{Hr}%
{c})^{2}} , \label{23}%
\end{equation}%
\begin{equation}
f_\mathrm{3}(r)=1-(\frac{Hr}{c})^{2} , \label{24}%
\end{equation}
which leads to the following expression:%
\begin{eqnarray}
&&M_\mathrm{u}(t)=4\pi\rho_\mathrm{0}(t)\left(  \frac{c}{H}\right) ^{3}\times\cr
&&\times\int_\mathrm{0}^{\xi_\mathrm{u}}\frac{\xi^{2}\mathrm{d}\xi}{\sqrt{1-\frac{2}{\pi\xi\xi
_\mathrm{u}^{2}}[\arcsin\xi-\xi\sqrt{1-\xi^{2}}]}\sqrt{1-\xi^{2}}} , \label{25}%
\end{eqnarray}
where the notations $\xi=Hr/c$ and $\xi_\mathrm{u}=HR_\mathrm{u}/c$ have been used.

It turns out that the above formula for the mass of the universe as function
of the world radius $R_\mathrm{u}$ in the limit $\xi_\mathrm{u}\rightarrow1$ is numerically
given by:%
\begin{equation}
lim_\mathrm{\xi_\mathrm{u}\rightarrow1}M_\mathrm{u}(t)=4\pi\rho_\mathrm{0}(t)R_\mathrm{u}^{3}\cdot1.6150446 , 
\label{26}%
\end{equation}
and using now the result from given in Equ. (\ref{14}), i.e. $\rho_\mathrm{0}%
(R_\mathrm{u})$ $=c^{2}/(2\pi^{2}GR_\mathrm{u}^{2})$, then yields the astonishing result:%
\begin{equation}
M_\mathrm{u}(t)=1.6150446 \frac{4\pi c^{2}R_\mathrm{u}^{3}}{2\pi^{2}GR_\mathrm{u}^{2}}\approx
\frac{c^{2}}{G}R_\mathrm{u} , \label{27}%
\end{equation}
revealing the fact that the above-defined mass of the universe $M_\mathrm{u}(t)$ in
the form it is conceived again scales with the corresponding radius of the universe.

On the other hand, turning things around, this also states that the radius of
the universe is about equal to the Schwarzschild radius of this mass
$M_\mathrm{u}(t)$ of the universe, since one finds:%
\begin{equation}
R_\mathrm{u}=\frac{1}{2}\frac{2GM_\mathrm{u}(t)}{c^{2}}\approx0.5\cdot R_\mathrm{s,u} . \label{28}%
\end{equation}

\section{Conclusions}

We have shown that with the use of an analogy to the inner stellar
Schwarzschild metric applied to the cosmic matter distribution one can arrive
at a reasonable definition of what could be called the mass $M_\mathrm{u}(t)$ of the
universe. Interestingly enough, this mass $M_\mathrm{u}(t)$ scales with the critical
radius $R_\mathrm{u}$ of the universe introduced by Equ.(\ref{10}) which was a
request already since the works by Mach (1983), Thirring (1918) (see also
Mashhoon et al. 1984, Barbour \& Pfister 1995, Wesson 1999, Hoyle 1990, 1992,
Jammer 2000 or Fahr \& Heyl 2006). Furthermore the finding that the average
density $\rho_\mathrm{0}$ of the universe turns out to be scaling with the reciprocal
of the square of the above defined critical radius $R_\mathrm{u}$ of the universe
just fulfills the request for an economical universe with vanishing total
energy as was discussed by Overduin \& Fahr (2003), Fahr (2004) and Fahr \&
Heyl (2006). In the near future we have plans to also consistently include
into the considerations the cosmic vacuum energy density to see what changes
might result from that for the above derived concepts.

An outstanding problem perhaps may still be to better understand the natural
philosophical semantics and the logical or physical implications of the above
given definitions of $M_\mathrm{u}$ and $R_\mathrm{u}$. One should perhaps notice that in
the above derived formulae no strict physical reason is presented why cosmic
matter density $\rho_\mathrm{0}$ scales with $R_\mathrm{u}^{-2}$. We only have derived above
a critical world radius $R_\mathrm{u}=R_\mathrm{u}(\rho_\mathrm{0})$ which is given as function of
the prevailing cosmic matter density $\rho_\mathrm{0}$ in a way such that the
relation $\rho_\mathrm{0}\sim R_\mathrm{u}^{-2}$ is fulfilled. Only when by some
cosmological evolution process not discussed here this density changes by its
value in time, making $\rho_\mathrm{0}=\rho_\mathrm{0}(t)$, then it turns out that the
accordingly changing critical radius $R_\mathrm{u}=R_\mathrm{u}(\rho_\mathrm{0}(t))$ also changes,
as if the relation $\rho_\mathrm{0}\sim R_\mathrm{u}^{-2}$ would be valid. The many
cosmological implications of that shall have to be discussed in more detail in
forthcoming considerations.


\begin{thebibliography}{}

\bibitem{}Barbour ,J.B.: 1995, in: J.B. Barbour, H. Pfister, {\it General relativity as a perfectly
Machian theory}, p. 214

\bibitem{}Barbour, J.,  Pfister, H.: 1995, in: '{\it Mach' principle: from Newton`s
bucket to quantum gravity'}, Birkh\"{a}user Verlag, Berlin

\bibitem{}Dirac, P.A.M.: 1937, Nature 139, 323

\bibitem{}Einstein, A.,  Straus, E.G.: 1945, Rev.Modern Phys. 17, 120

\bibitem{}Fahr, H.J.: 2004, in:  W.Loeffler,  P.Weingartner (eds.), {\it Knowledge and Belief -
Wissen und Glauben},, 26.th Int.Wittgenstein Symposium, \"{o}bv\&hpt, Wien

\bibitem{}Fahr, H.J.,  Heyl, M.: 2006, Astrophys.Space Sci., submitted

\bibitem{}Hoyle, F.: 1990, Astrophys. Space Sci. 168, 59   

\bibitem{}Hoyle, F.: 1992, Astrophys. Space Sci. 198, 195   

\bibitem{}Jammer, M.: 2000, in:  {\it  Concepts of Mass in contemporary Physics and
Philosophy}, Princeton University Press, Princeton

\bibitem{}Mach, E.: 1983, in: {\it Die Mechanik in ihrer Entwicklung,
historisch-kritisch dargestellt}, F.A.Brockhaus, Leipzig

\bibitem{}Mashhoon, B., Hehl, F.H., Theiss, D.S.: 1984, Gen.Rel.Grav.
16, 712  

\bibitem{}Overduin, J.,  Fahr, H.J.: 2003, Foundations of Physics Letters
16(2), 119  

\bibitem{}Rosen, N., Copperstock, F.I.: 1992, in: {\it The mass of a body in
general relativity}, Class. Quantum Gravity, 9, p.2657  

\bibitem{}Schuecking, E.: 1954, Zeitschrift f. Physik  137, 595

\bibitem{}Stephani, H.: 1988,  in:  {\it Allgemeine Relativit\"{a}tstheorie}, VEB
Deutscher Verlag der Wissenschaften, Berlin, p.113  

\bibitem{}Thirring, H.: 1918, Phys. Zeitschrift 19, 33

\bibitem{}Tolman, R.: 1934, in:  {\it Relativity, Thermodynamics and Cosmology},
Clarendon Press, Oxford, p.235

\bibitem{}Wesson, P.S.: 2000, Observatory 120, 59

\end{thebibliography}
\end{document}